\documentstyle[prl,aps,preprint,tighten,floats,epsf,rotate]{revtex}

\newbox\rotbox

\begin{document}



%
\preprint{\vbox{\rm\hfill TRI-PP-96-2\\}}
\title{Isospin Breaking in the Nucleon Isovector Axial Charge\\
        from QCD Sum Rules}
\author{Xuemin Jin}
\address{TRIUMF, 4004 Wesbrook Mall, \\Vancouver, 
British Columbia, Canada V6T 2A3\\}
%
%
\maketitle
\begin{abstract}
The isospin breaking in the nucleon isovector axial charge, $g_A^3$, 
are calculated within the external field QCD sum-rule approach. The
isospin violations arising from the difference in up and down current
quark masses and in up and down quark condensates are included;  
electromagnetic effects are not considered. We find 
$\delta g^3_A/g^3_A \approx (0.5-1.0)\times 10^{-2}$, where 
$\delta g^3_A = (g^3_A)_p + (g^3_A)_n$ and $ g^3_A = [(g^3_A)_p 
- (g^3_A)_n]/2$. Using the Goldberger-Treiman relation, we also obtain 
an estimate of the isospin breaking in the pion-nucleon coupling constant,
$(g_{pp\pi_0}-g_{nn\pi_0})/g_{NN\pi} \approx (2-7) \times 10^{-3}$. 
\end{abstract}
%
\newpage

The nucleon isovector axial charge (or the nucleon axial vector
coupling constant) is defined through the nucleon matrix element
of the isovector axial current at zero momentum transfer
\begin{equation}
\langle N | \overline{u} \gamma_\mu \gamma_5 u
- \overline{d} \gamma_\mu \gamma_5 d | N\rangle
= g^3_A \, \overline{U}(p)\, \gamma_\mu \gamma_5 \, U(p)\ ,
\label{ga-def}
\end{equation}
where $U(p)$ denotes the nucleon spinor. Assuming isospin symmetry, 
one finds $(g^3_A)_p = - (g^3_A)_n$, and the value of $ (g^3_A)_p = 
1.2573 \pm 0.0028$, extracted from the neutron beta decay,
has been quoted in the literature \cite{pdt94}.  In nature, the isospin 
symmetry is broken by the current quark mass difference as well as the 
electromagnetic interaction, and thus $(g^3_A)_p \neq - (g^3_A)_n$.

Previous studies of the nucleon isovector axial charge in the framework 
of external field QCD sum-rule method  have been made by various 
authors \cite{belyaev83,chiu85}. However, to our best knowledge, the 
isospin breaking effects have been ignored in these studies. The goal 
of this Letter is to examine the difference between $(g^3_A)_p$ and 
$(g^3_A)_n$ using the external field QCD sum-rule approach, which has been 
used in studying various nucleon matrix elements of bilinear quark 
operators \cite{belyaev83,chiu85,ioffe84,chiu86,belyaev85,kolesinchenko84,%
belyaev86,ioffe92,jin93,jin95a,jin95b}. The isospin violation is reflected 
in $m_u \neq m_d$ and the isospin breaking in the vacuum condensates. 
Electromagnetic effects will not be included.

Let us start from the correlation function of the nucleon interpolating
field in the presence of a {\it constant} external isovector axial vector
field $Z^\mu$
\begin{equation}
\Pi_Z(q) \equiv i\int d^4{x}e^{iq\cdot x}
\langle 0|{{\rm T}[\eta_N(x)\overline{\eta}_N(0)]}|0\rangle_Z\ ,
\label{corr}
\end{equation}
where $\eta_N$ is the nucleon interpolating field introduced in 
Ref.~\cite{ioffe81}
\begin{eqnarray}
\eta_p(x) & = & \epsilon_{abc}
\left[{u^T_a}(x)C\gamma_\mu u_b(x)\right]
\gamma_5\gamma^\mu d_c(x)\ ,
\label{etap}
\\*[7.2pt]
\eta_n(x) & = & \epsilon_{abc}
\left[{d^T_a}(x)C\gamma_\mu d_b(x)\right]
\gamma_5\gamma^\mu u_c(x)\ ,
\label{etan}
\end{eqnarray}
where $u_a(x)$ and $d_c(x)$ stand for the up and down quark fields,
$a,b$ and $c$ are the color indices, and $C=-C^T$ is the charge
conjugation operator. The subscript $Z$ in
Eq.~(\ref{corr}) denotes that we are evaluating the correlation
function in the presence of the external isovector axial vector field $Z_\mu$; the
correlator Eq.~(\ref{corr}) should be calculated with an additional
term
\begin{equation}
\Delta {\cal L}\equiv -Z_\mu \left[\overline{u} \gamma^\mu \gamma^5 u
- \overline{d} \gamma^\mu \gamma^5 d\right]\ ,
\label{dl}
\end{equation}
added to the usual QCD Lagrangian. The up and down quark fields then satisfy
the modified equations of motion:
\begin{eqnarray}
(i\rlap{\,/}D-m_u-\rlap{/}Z \gamma^5)u(x)&=& 0\ ,
\\*[7.2pt]
(i\rlap{\,/}D-m_d+\rlap{/}Z \gamma^5)d(x)&=& 0\ .
\label{eq-mo}
\end{eqnarray}

To first order in the external field, the correlation function can be 
written as 
\begin{equation}
\Pi_Z(q) = \Pi_0(q) + Z_\lambda \Pi^\lambda(q)\ ,
\label{corr-exp}
\end{equation}
where $\Pi_0(q)$ is the correlation function in the absence of the 
external field which gives rise to the usual mass sum rules. Here
we are interested in the linear response to the external field 
given by $\Pi^\lambda(q)$. The QCD sum rules for $\Pi^\lambda(q)$ 
differ from those for $\Pi_0(q)$. The phenomenological representation 
for $\Pi^\lambda(q)$ at the hadron level contains a double pole at the 
nucleon mass whose residue contains the matrix element of interest.
In addition there are single pole terms which arise from the transition 
matrix element between the ground state nucleon and excited states. 
These later contributions are not exponentially damped after Borel 
transformation relative to the double pole term and should be retained 
in a consistent analysis of the sum rules. On the theoretical side of the 
sum rules expressed in terms of an operator product expansion (OPE) the 
external field contributes in two different ways--by directly coupling to 
the quark fields in the nucleon current and by polarizing the QCD vacuum.

The linear response of the correlation function, $\Pi^\lambda(q)$,
has three distinct invariant structures \cite{belyaev83,chiu85}:
\begin{equation}
\Pi^\lambda = \Pi_1(q^2) q^\lambda \rlap{/}q \gamma^5 
+ \Pi_2(q^2) \gamma^\lambda \gamma^5 
+ \Pi_3(q^2) i q_\rho \sigma^{\lambda\rho} \gamma^5\ .
\label{resp-dec}
\end{equation}
So, one may derive three QCD sum rules from the three invariant
functions, $\Pi_1(q^2)$, $\Pi_2(q^2)$, and $\Pi_3(q^2)$, respectively.
In principle, the predictions based on these sum rules should be the
same. In practice, however, one has to truncate the OPE and use a simple 
phenomenological ansatz for the spectral density; thus these sum rules 
usually have different merits. In particular, some sum rules works better 
than the others. This pattern has been seen in various external field 
sum rules \cite{belyaev83,chiu85,ioffe84,chiu86,belyaev85,kolesinchenko84,%
belyaev86,ioffe92,jin93,jin95a,jin95b}. As discussed extensively in 
Ref.~\cite{chiu85}, this may be attributed to the different asymptotic 
behavior of various sum rules. The phenomenological side of the 
external field sum rules contains single pole terms arising from the 
transition between the ground state and the excited states, whose
contribution is {\it not} suppressed relative to the double pole term 
and thus contaminates the double pole contribution. The degree of this 
contamination may vary from one sum rule to another. The sum rule with 
smaller single pole contribution works better. We refer the reader to 
Refs.~\cite{belyaev83,chiu85,jin93,jin95a,jin95b} for more discussions 
on this point. As pointed out in Refs.~\cite{belyaev83,chiu85}, 
the sum rule obtained from $\Pi_1(q^2)$ is the most stable one for the 
problem under study. As such, we shall focus on this stable sum rule and 
disregard the sum rules based on $\Pi_2(q^2)$ and $\Pi_3(q^2)$. 

It is straightforward to obtain the external field sum rules following
the techniques given in the literature. To include the isospin violation 
effects, we retain the terms linear in current quark masses and isospin 
breaking in the condensates. Here we truncate the OPE at the same level 
as in the previous studies. The OPE result for $\Pi_1(q^2)$ in the proton 
case is given by
\begin{eqnarray}
\Pi_1(q^2) =& - & {1\over 16\pi^4}\, q^2\, \ln(-q^2)
- {1\over 16\pi^2}\, \langle {\alpha_s\over \pi} G^2 \rangle_0\, {1\over q^2}
 +  {1\over 3\pi^2}\, {\kappa \over q^2}
+ {4\over 3\pi^2}\, m_u\, \langle \overline{u} u\rangle_0\, {1\over q^2}
\nonumber
\\*[7.2pt]
& + & {20 \over 9}\, \langle \overline{u} u\rangle_0^2\, {1\over q^4}
- {2 \over 9}\,  m_u\, \langle \overline{u} u\rangle_0\, \chi\, {1\over q^4}
 -  {2\over 3}\, m_d\, \langle \overline{d} d\rangle_0\, \chi\, {1\over q^4}\ ,
\label{ope-p}
\end{eqnarray}
where $\langle \hat{O} \rangle_0 \equiv \langle 0|\hat{O}|0\rangle$, and
$\chi$ and $\kappa$ denote the linear response of condensates to the 
external field
\begin{eqnarray}
& &\langle 0|\overline{u}\gamma_\mu\gamma_5 u -
\overline{d}\gamma_\mu\gamma_5 d|0\rangle_Z = 2\, Z_\mu\, \chi\ ,
\\*[7.2pt]
& &\langle 0|\overline{u} \widetilde{G}_{\mu\nu} \gamma^\nu u
- \overline{d} \widetilde{G}_{\mu\nu} \gamma^\nu d |0\rangle_Z
= 2\, Z_\mu\, \kappa \ ,
\label{sus}
\end{eqnarray}
with $\widetilde{G}_{\mu\nu} = {1\over 2} \epsilon_{\mu\nu\lambda\rho} 
G^{\lambda\rho}$. Here we have omitted all the polynomials in $q^2$ 
which vanish under the Borel transformation, and neglected the responses 
of the corresponding isoscalar current to the external isovector field. 
The analogous result for the neutron is
\begin{eqnarray}
\Pi_1(q^2) =& + &{1\over 16\pi^4}\, q^2\, \ln(-q^2)
   + {1\over 16\pi^2}\, \langle {\alpha_s\over \pi} G^2 \rangle_0\, {1\over q^2}
       -  {1\over 3\pi^2}\, {\kappa \over q^2}
          - {4\over 3\pi^2\,} m_d\, \langle \overline{d} d\rangle_0\, {1\over q^2}
\nonumber
\\*[7.2pt]
& - & {20 \over 9}\, \langle \overline{d} d\rangle_0^2\, {1\over q^4}
  + {2 \over 9}\,  m_d\, \langle \overline{d} d\rangle_0\, \chi\, {1\over q^4}
   +  {2\over 3}\, m_u\, \langle \overline{u} u\rangle_0\, \chi\, {1\over q^4}\ .
\label{ope-n}
\end{eqnarray}

The resulting QCD sum rules can be written as
\begin{eqnarray}
& &M^4\, E_1\, L^{-4/9} + {b\over 4}\, L^{-4/9} 
   +{4\over 3}\, \widetilde{\kappa}\, L^{-68/81}
    +{16\over 3}\, \left(1 - {\gamma\over 2} 
      - {\delta m\over 2\hat{m}} \right)\,
        \hat{m}\, a L^{-4/9}
\nonumber
\\*[7.2pt]
& &\hspace*{0.75cm} + {20\over 9}\, \left(1 - \gamma\right)\, a^2\, L^{4/9}
     - {2\over 9}\, \left(1 -  {\gamma\over 2} 
       - {\delta m\over 2\hat{m}} \right)\,
           \hat{m}\, a\, \widetilde{\chi}\, {1\over M^2}\, L^{-4/9}
\nonumber
\\*[7.2pt]
& &\hspace*{1.5cm}
- {2\over 3}\, \left(1 +  {\gamma\over 2} 
      + {\delta m\over 2\hat{m}} \right)\,
         \hat{m}\, a \,\widetilde{\chi}\, {1\over M^2}\, L^{-4/9}
= \widetilde{\lambda}_p^2\, \left[ {(g^3_A)_p\over M^2} 
         + A_p\right] e^{-M_p^2/M^2}\ ,
\label{sum-p}
\end{eqnarray}
for the proton case and
\begin{eqnarray}
& &M^4\, E_1\, L^{-4/9} + {b\over 4}\, L^{-4/9} 
       +{4\over 3}\, \widetilde{\kappa}\, L^{-68/81}
          +{16\over 3}\, \left(1 + {\gamma\over 2} 
              + {\delta m\over 2\hat{m}} \right)\,
                 \,\hat{m}\, a L^{-4/9}
\nonumber
\\*[7.2pt]
& &\hspace*{0.75cm} + {20\over 9}\, \left(1 + \gamma\right)\, a^2\, L^{4/9}
        - {2\over 9}\, \left(1 +  {\gamma\over 2} 
              + {\delta m\over 2\hat{m}} \right)\,
                \hat{m}\, a\, \widetilde{\chi}\, {1\over M^2}\, L^{-4/9}
\nonumber
\\*[7.2pt]
& &\hspace*{1.5cm}
- {2\over 3}\, \left(1 -  {\gamma\over 2} - {\delta m\over 2\hat{m}} \right)\,
      \hat{m}\, a \,\widetilde{\chi}\, {1\over M^2}\, L^{-4/9}
= - \widetilde{\lambda}_n^2\, \left[ {(g^3_A)_n\over M^2} 
                     + A_n\right] e^{-M_n^2/M^2}\ ,
\label{sum-n}
\end{eqnarray}
for the neutron case, where we only keep the terms up to first order in isospin 
violation and have defined $a\equiv -(2\pi)^2 \left( \langle\overline{u}u\rangle_0
+ \langle\overline{d}d\rangle_0\right)/2$, $b\equiv (2\pi)^2 
\langle (\alpha_s/\pi) G^2\rangle_0$, $\hat{m} \equiv (m_u + m_d)/2$,
$\delta m= m_d - m_u$, $\gamma \equiv \langle\overline{d}d\rangle_0/
\langle\overline{u}u\rangle_0 - 1$, $\widetilde{\kappa} \equiv -(2\pi)^2 \kappa$,
$\widetilde{\chi} \equiv -(2\pi)^2 \chi$, and $\widetilde{\lambda}^2_{p(n)}
\equiv 32 \pi^4 \lambda_{p(n)}$, with $\langle 0|\eta_{p(n)}|p(n)\rangle
= \lambda_{p(n)} v_{p(n)}$ and $\overline{v}_{p(n)} v_{p(n)} = 2 M_{p(n)}$.
Here $A_p$ and $A_n$ are the phenomenological parameters that
represent the sum over the contributions from all off-diagonal transitions 
between the nucleon and excited states, and $E_1\equiv 1-e^{-s_0/M^2}\left(
{s_0\over M^2}+1\right)$, which accounts for the sum of the contributions involving 
excited states only, where $s_0$ is an effective continuum threshold. We have also 
taken into account the anomalous dimension of the various operators through the factor 
$L\equiv\ln(M^2/\Lambda_{\rm QCD}^2)/\ln(\mu^2/\Lambda_{\rm QCD}^2)$.  We take the 
renormalization scale $\mu$ and the QCD scale parameter $\Lambda_{\rm QCD}$ to be 
$500\,\text{MeV}$ and $150\,\text{MeV}$ \cite{ioffe81}, respectively. It is easy to 
see that the sum rules give rise to $(g^3_A)_p = - (g^3_A)_n$ when isospin violation is 
switched off.

To analyze the above sum rules and extract the quantities of interest, we adopt 
the numerical optimization procedures used in Refs.~\cite{jin93,jin95a,jin95b}.
The sum rules are sampled in the fiducial region of Borel $M^2$, where
the contributions from the high-dimensional condensates remain small and the 
continuum contribution is controllable. We choose $0.8\leq M^2\leq 1.4\, 
{\mbox{GeV}}^2$  which has been identified as the fiducial region for the nucleon 
mass sum rules \cite{ioffe84}. Here we adopt these boundaries as the maximal limits 
of applicability of the external field sum rules. The sum-rule predictions are obtained 
by minimizing the logarithmic measure
$\delta (M^2)={\mbox{ln}}[{\mbox{maximum}}\{{\mbox{LHS,RHS}}\}/
{\mbox{minimum}}\{{\mbox{LHS,RHS}}\}]$ averaged over $150$ points evenly spaced within 
the fiducial region of $M^2$, where LHS and RHS denote the left- and right-hand sides of 
the sum rules, respectively. Note that the coupling strengths $\lambda_{p(n)}^2$ 
also appear in the external field sum rules. Here we use the experimental values for 
the nucleon masses and extract $\lambda_{p(n)}^2$ from the nucleon mass sum rules 
given in Ref.~\cite{yang93} [Eqs.~(17) and (21)],
using the same optimization procedure as described above.
Since we neglect the electromagnetic interactions, we correct the neutron and
proton masses such that $M_n - M_p = 2.06\,\text{MeV}$, where
the central value for the electromagnetic contribution,
$(M_n - M_p)_{\rm el}  = -0.76\,\text{MeV}$ \cite{gasser82}, has been used.
We then extract $(g^3_A)_{p(n)}$, $A_{p(n)}$, and $(s_0)_{p(n)}$ from the sum 
rules Eqs.~(\ref{sum-p}--\ref{sum-n}).

\begin{figure}[t]
\begin{center}
\epsfysize=11.7truecm
\leavevmode
\setbox\rotbox=\vbox{\epsfbox{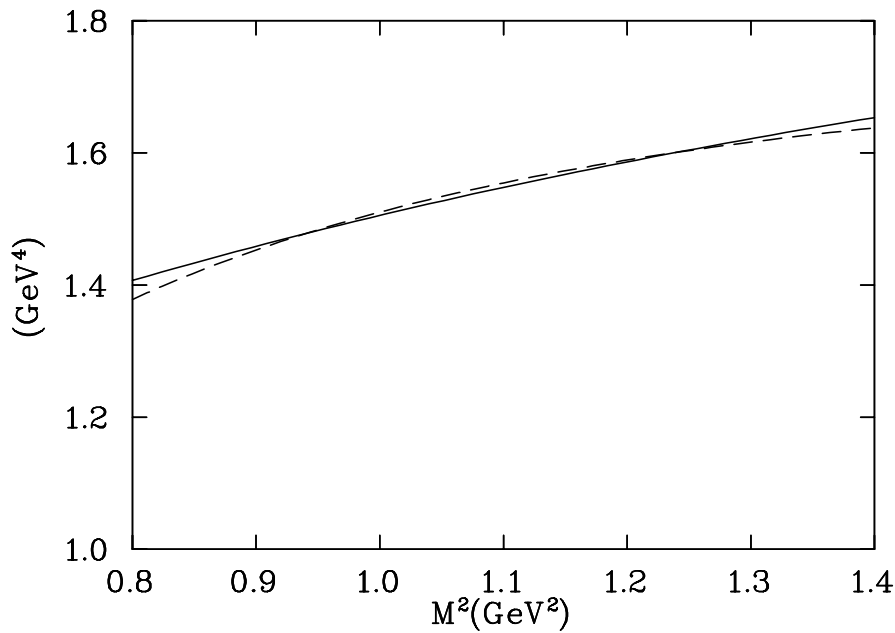}}\rotl\rotbox
\end{center}
\caption{The left-hand side (solid) and right-hand side (dashed)
 of Eq.~(\protect{\ref{sum-p}}) as functions of Borel $M^2$,
with $\gamma = -0.008$ and the optimized values for $(g^3_A)_p$,
$A_p$ and $(s_0)_p$.}
\label{fig-1}
\end{figure}

For vacuum condensates, we use $a=0.55\, {\mbox{GeV}}^3\, (m_u
+m_d\simeq 11.8{\mbox{MeV}})$ and $b=0.5\, {\mbox{GeV}}^4$ \cite{ioffe84}.
The parameters $\chi$ and $\kappa$ have been estimated previously.
Here we just quote the values, $\chi = - 2 f_\pi^2$ \cite{belyaev83,chiu85} and 
$\kappa \simeq -0.2\,\chi$ \cite{novikov84}, with $f_\pi = 93\,\text{MeV}$.
The quark mass difference $\delta m$ has been determined  by Gasser and Leutwyler, 
$\delta m /(m_u + m_d ) = 0.28 \pm 0.03$ \cite{gasser82}. The value of $\gamma$ 
has been estimated previously in various approaches \cite{paver87,pascual82,bagan84,%
dominguez85,dominguez87,narison89,adami91,adami93,eletsky93}. All of them, 
with the exception of Refs.~\cite{adami93,eletsky93}, indicate an interval 
of $-0.01 \le \gamma \le -0.006$. 
For these $\gamma$ values, we obtain
\begin{equation}
\delta g^3_A \simeq (0.5-1.0)\times 10^{-2}\ ,
\label{result}
\end{equation}
where $\delta g^3_A = (g^3_A)_p + (g^3_A)_n$ and $g^3_A = [(g^3_A)_p - (g^3_A)_n]/2$.
With smaller magnitude for $\gamma$, we get smaller values for $\delta g^3_A$.
To see how well the sum rules work, we plot the left- and right-hand sides
of the sum rule Eq.~(\ref{sum-p}) as functions of $M^2$ in Fig.~\ref{fig-1}, 
with $\gamma = -0.008$. One can see that the two sides have a good overlap.
This is typical for other values of $\gamma$ and for the neutron case.

As emphasized above as well as in the literature, the contribution
from the transition between the ground state nucleon and the excited 
states is not suppressed relative to the double pole term of interest.
This contribution is included through a single constant parameter, $A_{p(n)}$. 
This, as pointed out in Ref. \cite{ioffe95}, is an approximation. 
In principle, $A_{p(n)}$ should also be dependent on $M^2$. The impact
of approximating $A_{p(n)}$ as a constant on the extracted quantities
is expected to be small \cite{ioffe95}. Moreover, we have treated the 
continuum threshold $s^0_{p(n)}$ as a free parameter to be extracted
from the sum rules. This should partially account for the $M^2$
dependence of $A_{p(n)}$.

It is also worth pointing out that unlike the mass there are no experimental 
values for the couplings, $\lambda^2_{p(n)}$. One usually evaluates these
parameters from the mass sum rules by fixing the mass at the
experimental value.  This means that the uncertainties associated with 
$\lambda^2_{p(n)}$ will give rise to additional uncertainties in the 
determination of the nucleon matrix elements of various current, besides the 
uncertainties in the external field sum rules themselves. This is a general 
drawback of external field sum-rule approach and/or QCD sum-rule calculations
based on three point functions. Here we have not considered
the uncertainties associated with $\lambda^2_{p(n)}$.

From our calculation of the isospin breaking in the nucleon isovector
axial charge, we may estimate the isospin splitting in the pion-nucleon
coupling constants by invoking the Goldberger-Treiman relation. The pion
nucleon couplings are defined through the interactions
\begin{equation}
{\cal L}_{pp\pi_0} = g_{pp\pi_0} \overline{p} i\gamma_5 \pi_0 p\ ,
\hspace*{1cm}
{\cal L}_{nn\pi_0} = - g_{nn\pi_0} \overline{n} i\gamma_5 \pi_0 n\ .
\label{pion-n}
\end{equation}
Noth that both $g_{pp\pi_0}$ and $g_{nn\pi_0}$ are positive in this notation.
The Goldberger-Treiman relation then states 
\begin{equation}
(g^3_A)_p = g_{pp\pi_0} {f_\pi\over M_p}\ ,
\hspace*{1cm}
(g^3_A)_n = -g_{nn\pi_0} {f_\pi\over M_n}\ .
\label{gt}
\end{equation}
Using our results for $(g^3_A)_p$ and $(g^3_A)_n$, we find
\begin{equation}
{ g_{pp\pi_0} - g_{nn\pi_0} \over g_{NN\pi_0} }\approx (2 - 7)\times 10^{-3}\ ,
\label{pion-n-est}
\end{equation}
where $g_{NN\pi_0} = (g_{pp\pi_0} + g_{nn\pi_0})/2$. 
Since the Goldberger-Treiman relation only holds approximately and there
may be corrections to this relation which are originated from the isospin
breaking effects, our estimate here is only qualitative. Nevertheless,
the estimate given by Eq.~(\ref{pion-n-est}) is qualitatively compatible 
with the recent result obtained by Henley and Meissener \cite{henley96} 
from the QCD sum rules based on three point function, though the magnitude 
is somewhat smaller than that given in Ref.~\cite{henley96}. Our estimate here is also
consistent with those found in various models \cite{henley87,piekarewicz95,alaithan88,%
cao84}. On the other hand, Ref.~\cite{thomas81} gives a result which has 
a opposite sign.

In summary, we have calculated the isospin breaking in the nucleon 
isovector axial charge within the external field QCD sum-rule method. 
We included the isospin breaking effects due to the difference in 
current quark mass difference and in quark condensates, and neglected
the electromagnetic effects. We found a small isospin violation in
the nucleon isovector axial charge, $\delta g^3_A/g^3_A \approx (0.5-1.0)
\times 10^{-2}$. This, upon using the Goldberger-Treiman relation, 
leads to an estimate of the isospin breaking in the pion-nucleon coupling constant,
$(g_{pp\pi_0} - g_{nn\pi_0})/ g_{NN\pi_0} \approx (2 - 7)\times 10^{-3}$,
which is qualitatively consistent with previous studies.


\vspace*{1cm}
This work was supported by the Natural Sciences and Engineering
Research Council of Canada.


%
%

\end{document}